\documentclass[twocolumn,superscriptaddress,amsmath,amssymb,showpacs,prb]{revtex4-2}

\usepackage{graphicx}
\usepackage{color}
\usepackage{makecell}
\renewcommand{\phi}{\varphi}

\begin{document}

\title{Metallization and Spin Fluctuations in Cu-doped Lead Apatite }

\author{Yang Sun}
    \email{yangsun@xmu.edu.cn}
	\affiliation{Department of Physics, Xiamen University, Xiamen 361005, China}
\author{Kai-Ming Ho}
	\affiliation{Department of Physics, Iowa State University, Ames, IA 50011, USA}
\author{Vladimir Antropov}
    \email{antropov@ameslab.gov}
	\affiliation{Department of Physics, Iowa State University, Ames, IA 50011, USA}
    \affiliation{Ames National Laboratory, U.S. Department of Energy, Ames, IA 50011, USA}

\begin{abstract}
An electronic structure and magnetic properties analysis of the recently proposed Cu-doped lead apatite is performed. We show that electronic structures of differently Cu-substituted structures are characterized by localized molecular Cu-O bands at or near the Fermi level. The Cu substitutions can happen at both Pb1 and Pb2 sites, leading to metallic and semiconducting states differently. The electronic states in these bands are highly unstable magnetically and form clusters of rigidly ferromagnetically coupled magnetic moments on Cu and neighboring oxygen atoms with a total moment of about 1 $\mu_B$. The ground state of uniformly Cu-doped lead apatite appears to be magnetic and semiconducting. The non-uniform distribution of two Cu atoms at the nearest Pb2 sites leads to an antiferromagnetic semiconducting state with formation energy close to uniformly distributed Cu configurations. The inclusion of quantum spin fluctuations confirms the stability of magnetic Cu-O clusters. Our calculations revealed the absence of the long-range magnetic order between uniformly distributed Cu-O clusters, creating the spin glass type of system. 
\end{abstract}

\maketitle
\section{Introduction}
The recently discovered Cu-doped lead apatite (LK99) has attracted significant interest due to the so-claimed room-temperature superconductivity at ambient pressures by Lee et al. \cite{1,2}. So far,  the appearance of superconductivity in this compound is controversial in different experiments, and sometimes semi-conducting feature was observed \cite{3,4}. Density functional theory (DFT) calculations showed a flat band with Van Hove singularity (VHS) near the Fermi level for Cu substitution at specific Pb1 sites \cite{5,6}. The presence of such VHS can lead to many possible instabilities in electron, phonon, and spin fluctuations spectra. Thus, the nature of possible superconductivity in this system can result from several complex interactions, creating a large field of possible explanations. 

While there is no consensus on understanding unconventional superconductors, spin fluctuations (SF) have been identified in many high-Tc superconductors \cite{7} and Fe-based superconductors \cite{8} with a clear suggestion about their crucial influence on the mechanisms of unconventional superconductivity. In most cases, superconductivity appears near the magnetic quantum critical point, where magnetism has a pure itinerant character with no local moments involved. In such a case, quantum SFs (including spin zero-point motion) are strong and contribute to many observable physical properties. Therefore, the magnetic nature of LK99 and its origin can be crucial to the possibility of superconductivity in this system.

In this work, we use first-principles calculation with different DFT functionals and random-phase approximation (RPA) to study the Cu doping effect on the phase stability and electronic and magnetic prosperity of LK99. We aim to clarify the origin of metallization or semi-conductivity in this compound and examine competition between local moment and spin fluctuations.

\section{Methods}
Density functional theory (DFT) calculations were carried out using the projector augmented wave (PAW) method  \cite{9} implemented in the VASP code \cite{10,11}. PAW potentials with valence electronic configurations 3s$^2$3p$^3$, 6s$^2$5d$^{10}$6p$^2$, 3d$^{10}$4s$^1$, and 2s$^2$2p$^4$ were used for P, Pb, Cu, and N atoms, respectively. The structure relaxations were performed with generalized gradient approximation (GGA) parameterized by the Perdew-Burke-Ernzerhof formula (PBE)  \cite{12}. Electronic structures were studied with PBE, local density approximation (LDA) \cite{13}, and LDA+U \cite{14}. LDA+U calculations were performed using the simplified formulation of Dudarev et al. \cite{15}. A plane-wave basis was used with a kinetic energy cutoff of 520 eV. The convergence criteria were 10$^{-5}$ eV for the total energy and 0.01 eV/$\text{\AA}$ for the atomic force. The $\Gamma$-centered Monkhorst-Pack grid was adopted for Brillouin zone sampling with a spacing of $2\pi \times 0.033 ~\text{\AA}^{-1}$ for structure relaxation and $2\pi \times 0.017 ~\text{\AA}^{-1}$ for density of state calculations. 

The calculations of spin susceptibility and the exchange coupling parameters were based on our in-house electronic structure code's implemented full-potential linear augmented plane waves (FLAPW) method  \cite{16}. The spatial dependence of the susceptibility functions is represented using the mixed product basis set that consists of numerical functions inside the muffin-tin spheres and plane/dual-plane waves in the interstitial region. The details of the method are described in Ref.  \cite{16}.

\section{Results and discussion}

\subsection{Structure and stability}

The parent phase of LK99 (without Cu), Pb$_{10}$(PO$_4$)$_6$O, was determined in an apatite structure. A typical apatite mineral has a formula of Ca$_{10}$(PO$_4$)$_6$X$_2$, where X can be F, Cl, or OH in an oxidation state of -1. In Pb$_{10}$(PO$_4$)$_6$O, when the O$^{2-}$ ion occupies the halide site X, it leads to vacancies at other halide sites due to different oxidation states from the halide ion. Previous experiments determined that the O at the Pb$_{10}$(PO$_4$)$_6$O X site has a partial occupancy of 25$\%$ \cite{17}.

As shown in Fig.~\ref{fig:fig1}, the crystal structure of Pb$_{10}$(PO$_4$)$_6$O contains two types of Pb atoms. The Pb1 is in an independent layer and forms a hexagonal framework. The Pb2 is in the same layer as PO$_4$ and forms a triangular sublattice surrounding X sites (see Fig.~\ref{fig:fig1}(c)). To simulate this phase and maintain the neutral oxidation states, we position one O atom at the X site, noted as bridge-O in Fig.~\ref{fig:fig1}(d) (we will explain the name of “bridge” later). After structure relaxation, the bridge-O position slightly shifts from the X site and breaks the original symmetry from P6$_3$/m to P3, as shown in Fig.~\ref{fig:fig1}(d). The energy of the Pb$_{10}$(PO$_4$)$_6$O model is only 5 meV/atom above the Pb-P-O convex hull, which can be regarded as a stable phase at ambient conditions.

\begin{figure}[h]
\includegraphics[width=0.5\textwidth]{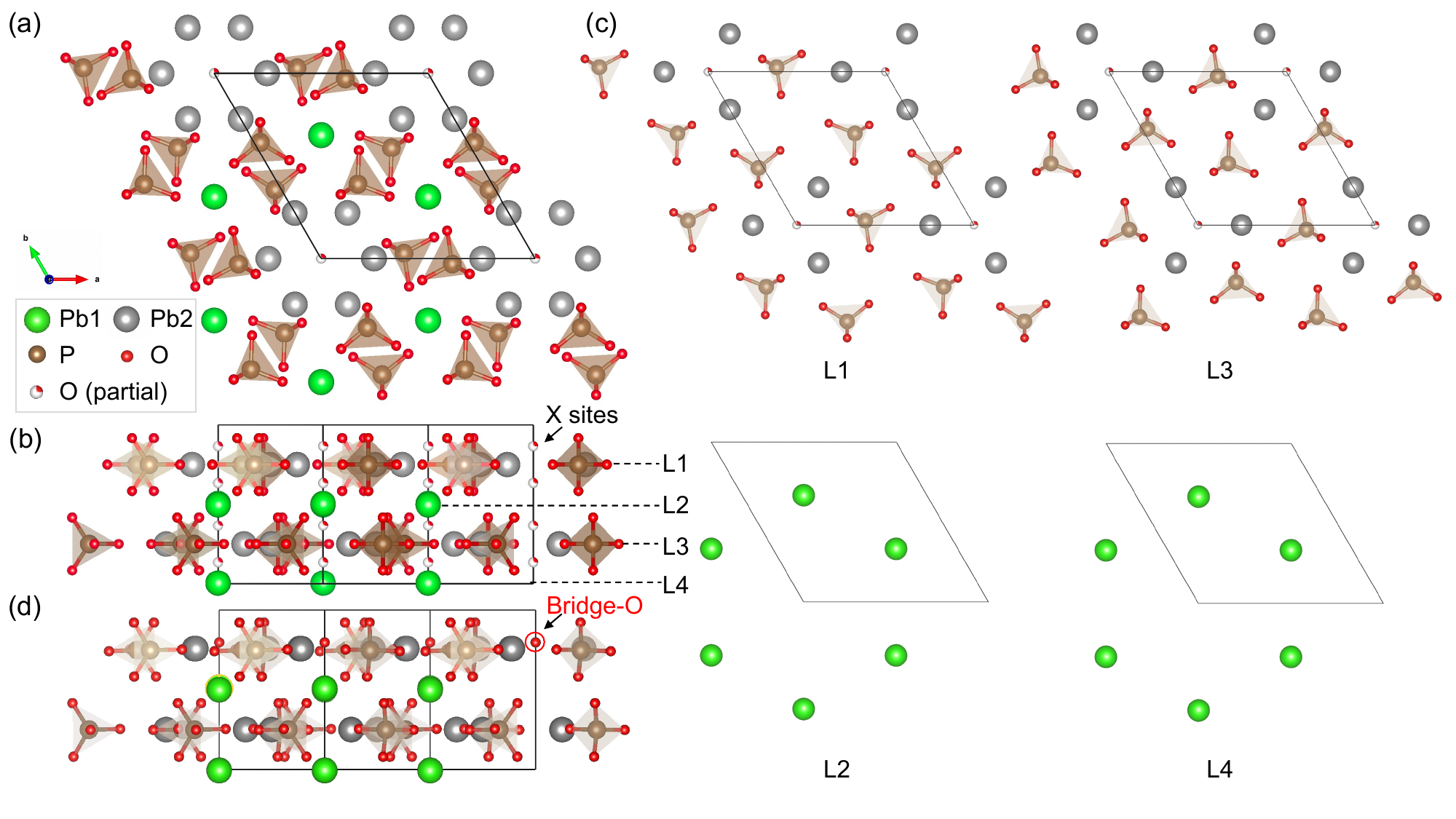}
\caption{\label{fig:fig1} Atomic structure of Pb$_{10}$(PO$_4$)$_6$O. (a) Top view. (b) Side view. (c) Motif in each layer. (d) Relaxed structure by DFT calculations.}
\end{figure}

Since Lee et al. reported a 9-10 at.$\%$ Cu doping at the Pb site  \cite{1},  we substitute one Pb by Cu in Pb$_{10}$(PO$_4$)$_6$O to examine the energetic stability. We keep the same experimental lattices and relax atomic positions. It results in four different Cu-doped structures: two inequivalent configurations with substitution at the Pb1 site and two inequivalent configurations with substitution at the Pb2 site. The inequivalence is due to the different relative positions between the Cu substitution site and the bridge-O atom. 

\begin{figure}[h]
\includegraphics[width=0.5\textwidth]{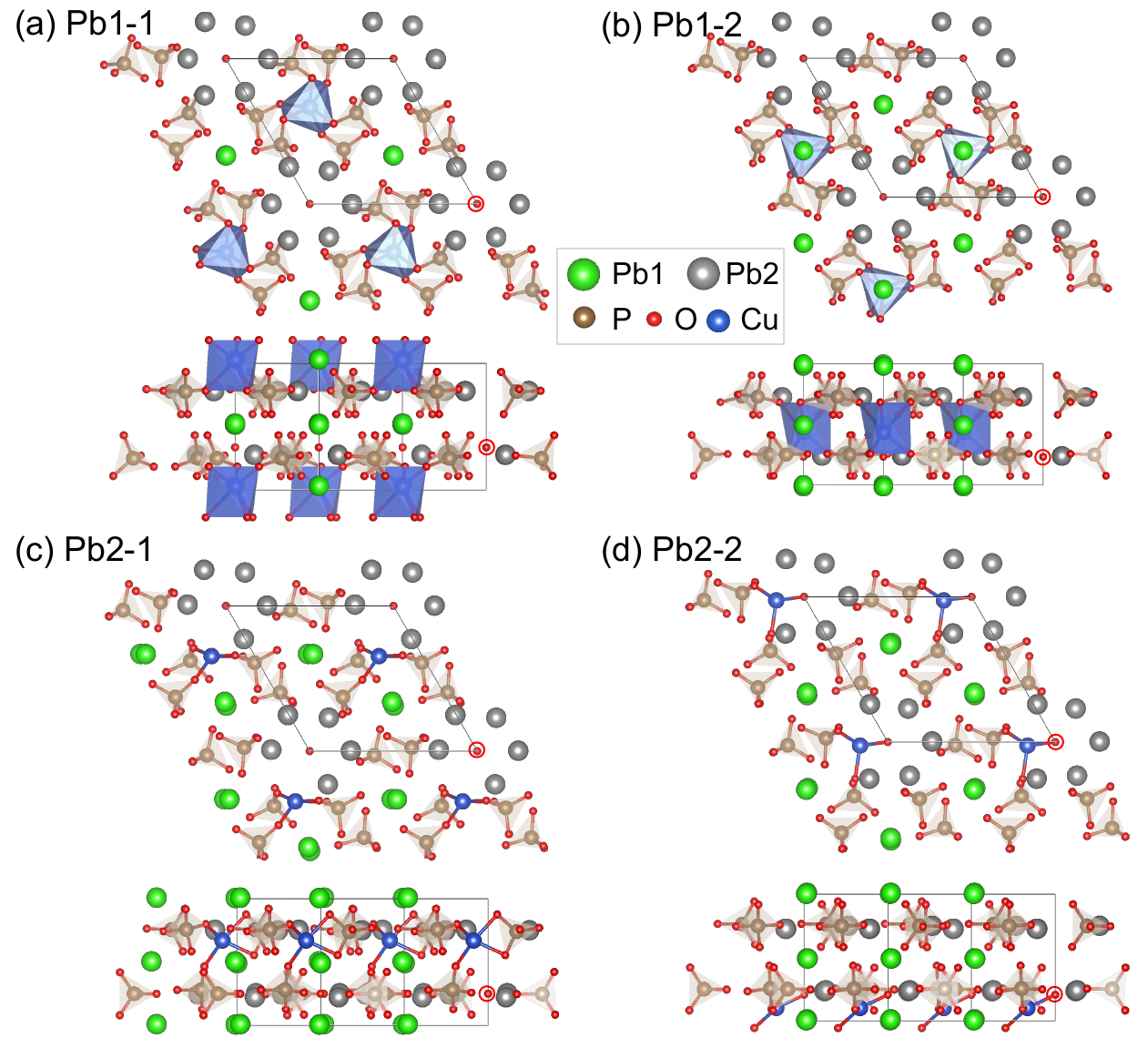}
\caption{\label{fig:fig2}  Configurations of Cu-doped Pb$_{10}$(PO$_4$)$_6$O. (a) and (b) show the Cu doping at the Pb1 site. (c) and (d) show the Cu doping at the Pb2 site. The upper and lower panels show top and side views, respectively. The bridge-O is circled in red.}
\end{figure}

Figure~\ref{fig:fig2} shows the relaxed configurations of Cu-doped in the lead apatite. For Pb1 site substitution, Cu forms a distorted triangular prism with surrounding O. This prism contains three short Cu-O bonds of $\sim$ 2.1 $\text{\AA}$ and three long Cu-O bonds of $\sim$ 2.5 $\text{\AA}$. Such distortion causes a tilt of surrounding PO4. For Pb2 substitution, the Cu forms a CuO4 polyhedron at the Pb2-1 site and CuO$_3$ at the Pb2-2 site. The bond length is 1.9-2.0 $\text{\AA}$. Such coordination is very different compared to PbO$_6$ polyhedrons at the same sites. Table 1 shows the spin-polarized calculations for the energetics difference among these structures. Cu substitutions at Pb1 sites generally have higher energy than those at Pb2 sites. The energy difference is insignificant, considering the large number of atoms affected in the cell. If no special treatment were performed in experiments, one would expect the Cu to substitute both Pb1 and Pb2 sites in the doped lead-apatite structure with a preference for the Pb2 site. Charge analysis suggests the most significant charge differences among these configurations come from the $s$ and $p$ orbitals of the Cu atom. The bridge-O atom contributes the second most significant charge difference among different structures. While this atom is quite far from the Cu, it is affected by the Friedel oscillations caused by the Cu substitution. This O atom participates in the energy transfer between Cu atoms and between cells. Therefore, we name it a “bridge” atom. The stability of the structure largely relies on these charge states. Table 1 also shows that LDA and PBE calculations lead to a qualitatively similar energy difference among different structures, suggesting that the stability analysis is robust. 

\begin{table}[h]
\caption{\label{table:tab1} The energy difference of Cu substitution at different Pb sites. $\Delta$E is the total energy difference for 41 atoms in Pb$_{10}$(PO$_4$)$_6$O.}
\centering
\begin{tabular} {c | c | c}
\hline
\hline
System       & $\Delta$E by PBE (eV/f.u.) & $\Delta$E by LDA (eV/f.u.)      \\
\hline
Pb1-1	&0.33 & 0.74 \\
\hline
Pb1-2	&0.78	&1.03 \\
\hline
Pb2-1	&0.29	&0.33\\
\hline
Pb2-2	&0.0	&0.0\\
\hline
\hline
\end{tabular}
\end{table}

We also consider the addition of a pair of Cu atoms with totally 45 possible configurations in Pb$_8$Cu$_2$(PO$_4$)$_6$O. PBE calculations show that the lowest energy configuration is the one with two Cu atoms substituting Pb2-2 sites (see Supplementary Fig. S3a). This leads to a peculiar  antiferromagnetic state, where two Cu atoms show non-equivalent magnetic moments, 0.43 $\mu_B$ and -0.38 $\mu_B$, respectively. While the magnetic moments on Cu atoms (mostly of $d$-character) are not exactly equal, $s+p$ contributions to the magnetic moment from neighboring atoms completely compensate for non-zero Cu's d-moment. Individual magnetic moments on Cu atoms in such antiferromagnetic Cu dimer appears to be somewhat weaker relative to a single Cu substitution. One can expect that the magnetism can be further weakened with the increase of local Cu concentration.  The DOS of this phase also shows the presence of Cu and O states near the Fermi level, similar to the DOS of the single Cu substitution at the Pb2-2 site (see Supplementary Fig. S3b). The formation energy difference between one and two Cu substitutions, given by $\Delta E=E( \text{Pb}_9 \text{Cu}(\text{PO}_4 )_6 \text{O})-0.5*[E(\text{Pb}_8 \text{Cu}_2 (\text{PO}_4 )_6 \text{O})+E(\text{Pb}_{10} (\text{PO}_4 )_6 \text{O})]$, is only around 1 meV/atom, where $E(\text{Pb}_9 \text{Cu}(\text{PO}_4 )_6 \text{O})$ is the ground-state energy of single Cu substitution at the Pb2-2 site. Such a small energy difference indicates the Cu substitution does not exhibit preferable ordering and the effects of clusterization can be expected in this system.

\begin{figure}[h]
\includegraphics[width=0.48\textwidth]{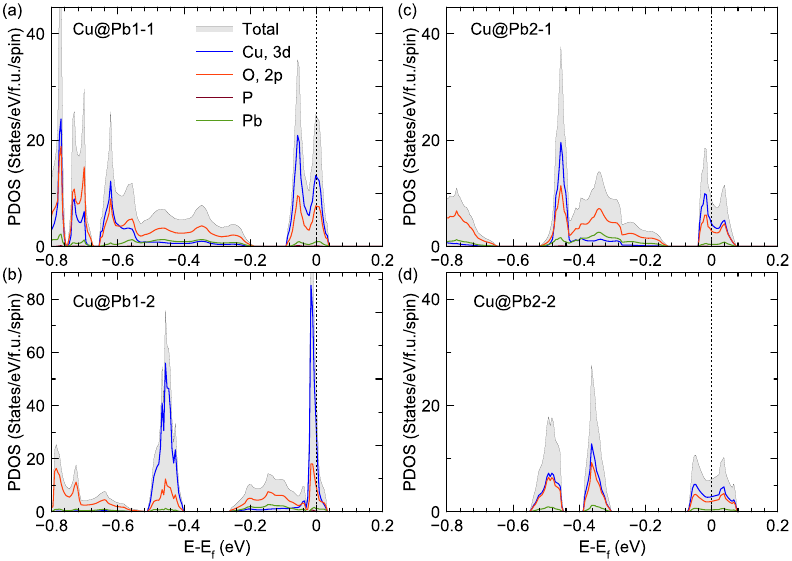}
\caption{\label{fig:fig3}  Non-magnetic density of states of Pb$_9$Cu(PO$_4$)$_6$O.}
\end{figure}

\subsection{Electronic structure}
Without Cu substitution, Pb$_{10}$(PO$_4$)$_6$O is a non-magnetic insulator with a band gap of $\sim$ 2.7 eV. We first consider the effect of Cu substitution on the non-magnetic electronic structures. Figure~\ref{fig:fig3} shows the non-magnetic density of states for the Pb$_9$Cu(PO$_4$)$_6$O. With Cu substitution, all four structures generated by different Cu substitution are metallic, with many states at the Fermi level, mainly contributed by Cu and O. Cu-$3d$ and O-$2p$ show a strong hybridization and form highly localized states (VHS) near the Fermi level. These VHSs are similar to isolated molecular states and have a narrow bandwidth of $\sim$ 0.15 eV. As we show below, the effective SF energy in this material has a similar strength, making charge and spin fluctuations competitive.

With such large states at the Fermi level, one would expect the appearance of local magnetic moment and/or ferromagnetism. Our non-magnetic spin susceptibility calculations revealed that the enhancement factor  $1-\hat{I}\hat{\chi}_0$ is negative on Cu and some oxygen modes, indicating several magnetic instabilities. Direct calculations revealed that the criterion of the local moment  \cite{18} is well fulfilled on the Cu atom in all considered configurations ($I$=0.52 eV). On oxygen atoms, the local moment criterion is not satisfied; however, the Stoner criterion is fulfilled, and the itinerant magnetism appears due to susceptibility contributed by Cu atoms. Our consecutive spin-polarized calculations confirmed that all four structures have magnetic ground states with a total magnetic moment of $\sim$ 1.0 $\mu_B$/f.u.

\begin{figure}[h]
\includegraphics[width=0.48\textwidth]{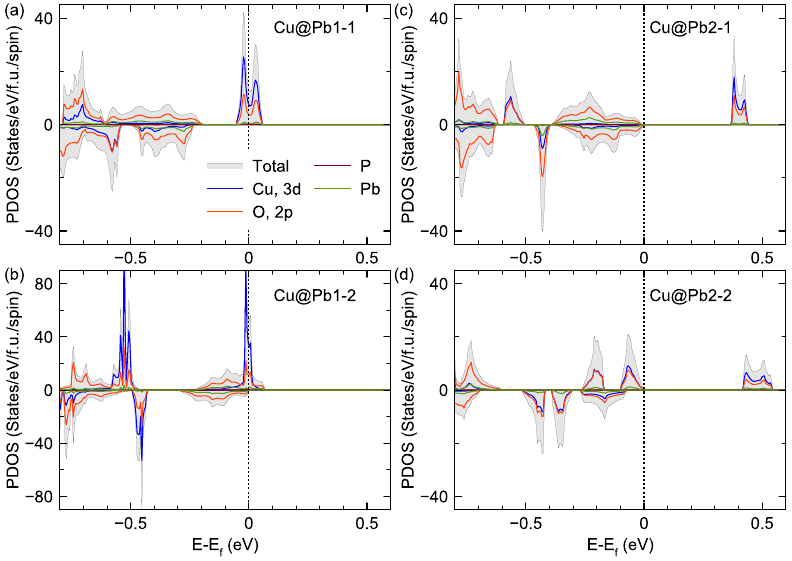}
\caption{\label{fig:fig4}  Spin-polarized density of states of Pb$_9$Cu(PO$_4$)$_6$O configurations with different Cu-doped configurations.}
\end{figure}

The magnetic instability induced by strong VHS leads to the change of metallization. As shown in Fig.~\ref{fig:fig4}, for Cu-substituted Pb1 phases, both configurations still maintain the metallic states. However, for Cu-substituted Pb2 phases, the magnetism causes a band mechanism of metal-insulator transition. It opens a semiconducting gap of $\sim$ 0.4 eV in a ferromagnetic state. Therefore, in the doped phase, the Cu replacement of the Pb2 atom can strongly affect the metallization. Qualitatively similar results are obtained with LDA and LDA+U calculations (Supplementary Materials Fig. S1 and S2).

\begin{figure}[h]
\includegraphics[width=0.38\textwidth]{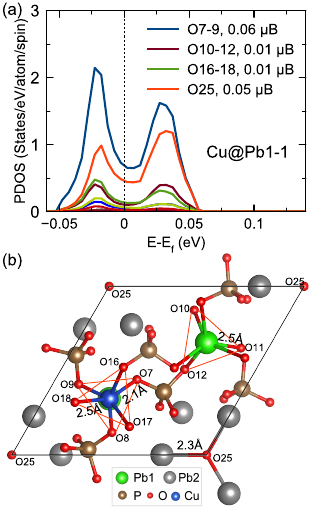}
\caption{\label{fig:fig5}   (a) Partial density of states of O atoms in Pb$_9$Cu(PO$_4$)$_6$O with Cu at Pb1-1 site. The legends indicate the partial magnetic moment on each O atom. (b) The atom positions and bond length.}
\end{figure}

To understand the metallic and magnetic interaction between Cu and O, we focus on the Cu@Pb1-1 doping configuration. The total magnetic moment of this phase is $\sim$ 1.0 $\mu_B$/f.u. with Cu atoms partial moment of 0.6 $\mu_B$, while neighboring O atoms contribute $\sim$ 0.3 $\mu_B$. Figure~\ref{fig:fig5}(a) shows all O atoms' partial DOS near the Fermi level. It indicates four groups of O atoms contributing to the DOS and magnetic moments: O7-9 for 0.06 $\mu_B$/atom, O10-12, and O16-18 for 0.01 $\mu_B$/atom, and O25 (i.e., the bridge-O atom) for 0.05 $\mu_B$/atom. As shown in Fig.~\ref{fig:fig5}(b), O7-9 is the nearest neighbor of Cu atoms with Cu-O bond length of 2.1Å. O10-12 atoms are the second nearest neighbors of Cu with a bond length of 2.5Å. O16-18 is much further and only bonded to Pb. The bridge-O atom interacts with several Cu atoms in different primitive cells, which results in its FM polarization (0.05 $\mu_B$). This polarization of bridge-O disappears if we consider the AFM order between Cu atoms in the different cells. Thus, the calculations of the exchange coupling between Cu atoms strongly depend on Cu spin orientations. In this case, the total energy differences would include the energy of bridge-O magnetic moment formation, making such an approach meaningless. To estimate the Heisenberg model ($H=-\sum_{i j} J_{i j} \boldsymbol{M}_j \boldsymbol{M}_j$) parameters, we used the traditional spin-polarized RKKY approach in the form suggested in Ref. \cite{19},

\begin{equation}
J_{i j}=\left(\chi^{-1}\right)_{i j} \approx-\Delta_i \chi_{i j} \Delta_j
\label{eq1},
\end{equation}
where $\chi_ij$ is a pairwise element of transversal ‘bare’ spin susceptibility, and the $\Delta_i$ is the on-site element of its inverse. The essential exchange interactions in the considered FM structure appear between Cu and O7 of 62 meV (same for O8 and O9), bridge-O of 44 meV, and nearest Cu atoms (out of the plane is 2.3 meV and in the plane is 0.7 meV), and all are positive for FM configurations. The coupling of the local moment on Cu with the nearest neighbor oxygen atoms is the strongest. Cu atom creates FM polarization of the surrounding oxygen atoms (O7-9), forming a relatively rigid spin cluster CuO$_3$. Cu clusters weakly interact with each other through bridge-O, creating something like the spin glass of such clusters. It happens at this concentration of Cu atoms; randomness in Cu positioning may create different spin clusters, while increasing Cu concentration would likely destroy this magnetic state. Similar magnetic interactions between Cu and the nearest neighbor O atoms and bridge-O are found for the Pb1-2 configuration of Cu substitution.

\subsection{Spin fluctuations addition}
LDA-based methods traditionally overestimate the magnetism for weakly magnetic systems due to the neglection of quantum SF at T=0K (spin zero-point motion)  \cite{20}. To study the influence of SF on the magnetic stability of considered above magnetic configurations, we first obtain the SF energy (absent in LDA) using the RPA technique  \cite{15,18}. Then, we calculate a corresponding SF contribution to the spin susceptibility and obtain the renormalized Stoner criterion for magnetic instability. The following RPA expression for the SF contribution to the total energy has been used: 

\begin{widetext}

\onecolumngrid

\begin{equation}
E_{S F}=\lim _{T \rightarrow 0} F_{S F}=2 \hbar \lim _{T \rightarrow 0} \sum_v \int_0^I d I \int_{\Omega_{B Z}} d \boldsymbol{q} \int_{-\infty}^{\infty} d \omega\left(\operatorname{Im} \chi_v(\boldsymbol{q}, \omega)-\operatorname{Im} \chi_{v 0}(\boldsymbol{q}, \omega)\right) \operatorname{coth}(\beta \omega / 2)
\label{eq2},
\end{equation}
\end{widetext}
\twocolumngrid
where $\chi$ and $\chi_0$ are enhanced and bare transversal spin susceptibility of LDA. Computational details of our linear response method can be found in Ref. \cite{16}. The transversal part of SF correction to the inverse magnetic susceptibility is obtained directly as  

\begin{equation}
\chi_{S F}^{-1}=\frac{1}{2 M} \frac{\partial}{\partial M}\left(\int_0^I d I\left(\left\langle M^2\right\rangle-\left\langle M_0^2\right\rangle\right)\right)
\label{eq3}.
\end{equation}

Here $⟨M^2⟩$  is a square mean of magnetic moment obtained with enhanced and nonenhanced susceptibility, correspondingly. The SF renormalized Stoner criterion then is written as

\begin{equation}
\left(\chi_{\mathrm{SF}}^{-1}+\chi^{-1}\right) \chi_0=1-I^* \chi_0>0
\label{eq4},
\end{equation}
where $I^*$ is a renormalized Stoner parameter, which includes itinerant SF beyond the usual static LDA/GGA. Moriya first noticed such renormalization of the interaction parameter \cite{20} that was studied much later using electronic structure calculations in \cite{22} for several magnetic 3d metals. All these studies have been performed using ad-hoc low-frequency and long-wavelength approximations. However, as we have shown recently (see \cite{19,22}), this approximation may not be reliable in realistic metallic systems. The above approach considers SF at all $q$ vectors and frequencies.

Our results indicated that the SF renormalization of the effective Stoner parameter in LK99 is a 12-17$\%$ decrease of the “bare” LDA value (0.5 eV) on Cu sites. With such renormalization, the criteria of the local moment on the Pb2-2 site is close to 1 and marginally not fulfilled anymore. This situation of marginal stability can be resolved only by performing self-consistent calculations. In contrast, the other three configurations we considered are still very unstable, and a strong local moment on the Cu atom should be formed there. 

The direct estimation of SF influence on magnetic moments was performed using the direct addition of non-local "spin fluctuation” potential to the self-consistent LDA magnetic state of LK99 in the form,

\begin{equation}
V_{S F}\left(\boldsymbol{r}, \boldsymbol{r}^{\prime}\right)=I(\boldsymbol{r})\left[\chi\left(\boldsymbol{r}, \boldsymbol{r}^{\prime}\right)-\chi_0\left(\boldsymbol{r}, \boldsymbol{r}^{\prime}\right)\right] I\left(\boldsymbol{r}^{\prime}\right)
\label{eq5}.
\end{equation}

The density functional self-consistency would require dealing with double-counting terms (the amount of quantum spin fluctuations in the local density functional is unknown), introducing a constraining spin field, and the original state losing its Kohn-Sham nature. Due to these factors, we consider here only a one-shot calculation of this correction for Cu-substituted Pb2-1 and Pb2-2 systems. These calculations revealed that adding VSF for both magnetic configurations weakens the magnetic moment of Cu atoms (by 0.09  $\mu_B$ and 0.12  $\mu_B$). Still, the magnetic moment on Cu and polarization on the nearest neighboring oxygen atoms remain. However, adding such potential also leads to a metallic state on the Pb2-2 configuration, while the Pb2-1 configuration remains semiconducting. We associate such differences with a strong presence of Cu states near the Fermi level in the Pb2-2 configuration. 

In contrast, these states practically disappear for the Pb2-1 configuration (see Fig.~\ref{fig:fig4}c, d). SF on oxygen sites can also be significant. Still, due to the induced nature of the polarization on these atoms, this magnetization is described correctly in LDA after Cu atom renormalization. Our calculations suggested that Cu atoms are more likely to go to Pb2 sites. If we assume, as described above, that Cu atoms occupy both Pb2 sites, we could have the coexistence of such Cu-rich and the original insulating Cu-less phases. The amount of each phase would strongly depend on the material's synthesis conditions.

While many different SF with q-vectors around the Fermi level contribute to this suppression of magnetic moment, spin wave contribution looks dominating in the frequency range up to 0.25 eV. Despite this energy range of SF seems accessible for neutron scattering experiments, the experimental verification can be problematic. Our estimation is made for the FM structure, but we cannot claim the stability of FM or AFM orders between Cu atoms. The strength of obtained SF is significant to eliminate the stability of such magnetic orderings but not strong enough to “kill’ local moment related to the CuO$_3$ cluster. This local moment survives in our RPA treatment of spin fluctuations. The system overall can be characterized as a system with ferromagnetic CuO$_3$ clusters with no order between them, similar to spin-glass behavior already at T=0K. We estimate the average SF energy as 0.12 eV, comparable with the effective Cu-O bandwidth at the Fermi level. Thus, these two scales can compete and interact.

\begin{figure}[h]
\includegraphics[width=0.4\textwidth]{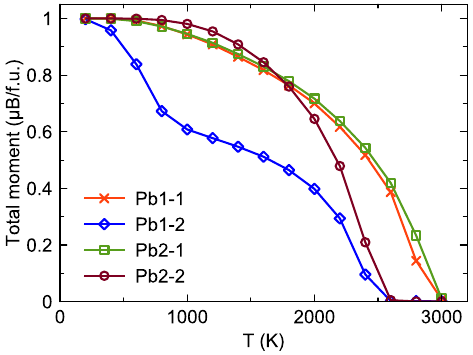}
\caption{\label{fig:fig6}  Magnetic moment as a function of electronic temperature for Pb$_9$Cu(PO$_4$)$_6$O.}
\end{figure}

To further check the nature of magnetic states in these systems, we studied the temperature dependence of magnetization and individual atomic moments in the band Stoner model  \cite{23}. In this model, the Curie temperature is determined by the Stoner parameter instead of the interactions among the localized spins, as in the Heisenberg model. In Fe, Co, and Ni, such a model strongly overestimates the experimental Curie temperature \cite{23}. Since that, such band description has been used only for weak magnetic systems. Nevertheless, it is methodically helpful to check how far our systems are from ferromagnetic Ni, where the magnetic moment is similar to the local moment on Cu (0.6 $\mu_B$) in our system. In Fig.~\ref{fig:fig6}, the Curie temperature of the Stoner model in our magnetic configurations appears like that of Ni (2500K - 3000K). This indicates that the moment on Cu is relatively localized. The induced magnetization on O atoms follows Cu moment behavior at all temperatures except the Pb1-2 configuration (see Fig. S4). These results also support the locality of Cu magnetic moment and magnetic stability of ferromagnetic Cu-O clusters.

\section{Conclusion}
In summary, we applied first-principles calculations to the Pb$_9$Cu(PO$_4$)$_6$O model of LK99. DFT-based calculations suggest that Cu substitutions at the Pb2 sites show better stability than those at the Pb1 sites, while their energies are close to each other. The stability largely depends on the charge state of Cu atoms. The bridge-O atom at the halide site of apatite participates in the energy transfer and is strongly affected by Cu's charge and spin states. For Pb$_8$Cu$_2$(PO$_4$)$_6$O, the ground-state configuration is two Cu atoms substituted nearest Pb2-2 sites and forms an antiferromagnetic semiconducting state. Such non-uniform distribution of Cu substitutions has very close formation energy with the uniformly distributed Cu configurations, suggesting that Cu substitution does not exhibit specific ordering. The electronic structure of Pb$_9$Cu(PO$_4$)$_6$O is characterized by localized, molecular-like Cu-O bands at or near the Fermi level. Non-magnetic calculations reveal that metallization appears in all Cu-substituted structures, characterized by highly localized states with a narrow bandwidth of $\sim$ 0.15 eV. Forming such VHS at the Fermi level can lead to numerous instabilities of a different nature in the electronic, lattice, or magnetic subsystems. While it is impossible to identify which instability would be the strongest, we analyze the magnetic instabilities in this work. With the introduction of magnetism, Cu substitution at the Pb1 site results in metallization, while Cu substitution at the Pb2 site generates a semiconductor with a band gap of $\sim$ 0.4 eV. The local moment on Cu predominantly couples with its nearest oxygen neighbor atoms, forming rigid CuO$_3$ spin clusters. Cu also weakly interacts with the bridge-O atom, creating a spin-glass-like interaction between CuO$_3$ clusters. The RPA calculation suggests that including spin fluctuations while weakening magnetism cannot eradicate the local moment associated with CuO$_3$ clusters for Pb$_9$Cu(PO$_4$)$_6$O. The Stoner temperature of the compound is estimated to be around 2500-3000  K, further supporting the localized nature of the Cu magnetic moment and the relatively strong magnetic stability of the ferromagnetic Cu-O clusters. In addition, spin fluctuations can change the semiconducting state, making the ground state of the Pb2-2 configuration metallic. Different Cu configurations could lead to the coexistence of metallic and insulating (semiconducting) phases. Our calculations also support another opportunity: the coexistence of the metallic phase with Cu on the Pb2-2 sites and the original insulating copper-less phase. Experimentally, the crystal phase has yet to be identified clearly, and the analogous presence of both metallic and semiconducting phases looks likely.  Computationally, in the density functional method, the uniformly doped Pb$_9$Cu(PO$_4$)$_6$O compound is an insulating (semiconducting) system with ferromagnetic Cu-O clusters that exhibit very little or no magnetic order, similar to a spin-glass behavior. Clustering effects could lead to the appearance of antiferromagnetism between Cu atoms.



\bibliographystyle{apsrev4-2}

\clearpage

\newpage



\onecolumngrid
\setcounter{equation}{0}
\setcounter{figure}{0}
\setcounter{table}{0}
\makeatletter
\renewcommand{\theequation}{S\arabic{equation}}
\renewcommand{\thefigure}{S\arabic{figure}}
\renewcommand{\bibnumfmt}[1]{[S#1]}
\renewcommand{\citenumfont}[1]{S#1}
\renewcommand{\thesection}{S\arabic{section}}

\setcounter{section}{0}

\begin{figure}
\includegraphics[width=0.7\textwidth]{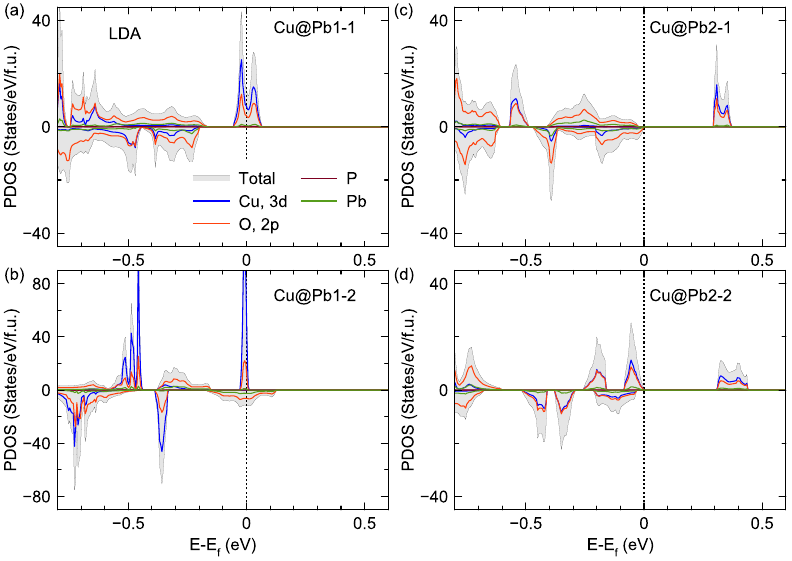}
\caption{\label{fig:fig1} Spin-polarized density of states of Pb$_9$Cu(PO$_4$)$_6$O with LDA calculations.}
\end{figure}

\begin{figure}
\includegraphics[width=0.7\textwidth]{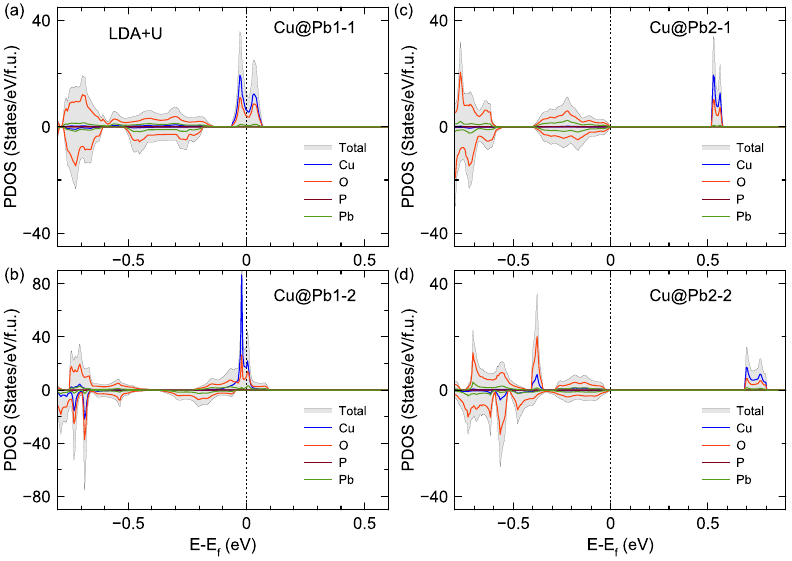}
\caption{\label{fig:fig2} Spin-polarized density of states of Pb$_9$Cu(PO$_4$)$_6$O with LDA+U calculations. The Hubbard parameter U is 4 eV.}
\end{figure}

\begin{figure}
\includegraphics[width=0.6\textwidth]{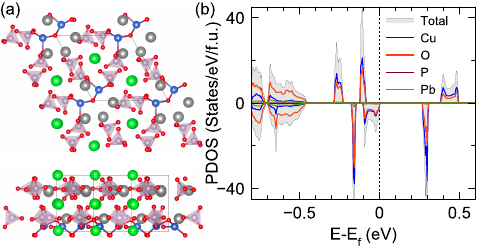}
\caption{\label{fig:fig3} (a) Lowest-energy configuration for two Cu substitutions in Pb$_8$Cu$_2$(PO$_4$)$_6$O. Both two Cu is at Pb2-2 sites. (b) Total density of state of Pb$_8$Cu$_2$(PO$_4$)$_6$O with two Cu at Pb2-2 sites.}
\end{figure}

\begin{figure}
\includegraphics[width=0.4\textwidth]{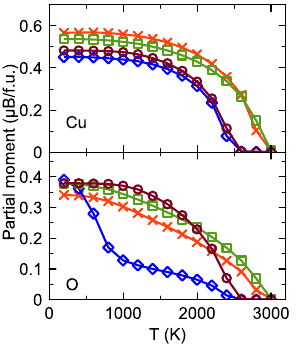}
\caption{\label{fig:fig4} Partial magnetic moment as a function of electronic temperature for Pb$_9$Cu(PO$_4$)$_6$O.}
\end{figure}

\end{document}